\documentclass[sigconf]{acmart}

\usepackage{adjustbox}
\usepackage{booktabs}
\usepackage{multirow}
\usepackage{xcolor}
\usepackage{float}

\AtBeginDocument{%
  }

\setcopyright{acmlicensed}
\copyrightyear{2026}
\acmYear{2026}
\acmDOI{XXXXXXX.XXXXXXX}

\acmConference[CONF '26]{Conference Name}{Month, 2026}{City, Country}

\begin{document}

\title{It's Not About Whom You Train: An Analysis of Corporate Education in Software Engineering}

\author{Rodrigo Siqueira}
\email{rms9@cesar.school}
\affiliation{%
  \institution{CESAR School}
  \city{Recife}
  \state{PE}
  \country{Brazil}
}

\author{Danilo Monteiro Ribeiro}
\email{dmr@cesar.school}
\affiliation{
  \institution{CESAR School}
  \city{Recife}
  \state{PE}
  \country{Brazil}
}

\renewcommand{\shortauthors}{Siqueira et al.}

% ============================================================
\begin{abstract}
\textbf{Context:} Corporate education is a strategic investment in the software industry, but little is known about how different professional profiles perceive these initiatives.
\textbf{Objective:} To investigate whether sociodemographic and professional variables influence the perception of quality and effectiveness of corporate training in Software Engineering (SE).
\textbf{Method:} Non-parametric significance tests were applied to data from a \textit{survey} with 282 Brazilian professionals, crossing 27 perception items with 9 sociodemographic variables (gender, age, education level, state, experience, professional level, company size, area of work, and nature of participation), totaling 243 combinations.
\textbf{Results:} Of the 243 combinations tested, only 35 showed statistical significance. Training mandatoriness was the dominant factor, affecting 24 of 27 items. Length of experience revealed a non-linear descriptive pattern with a low-engagement zone between 3 and 6 years. Differences by area of work indicated an expressive gap in \textit{soft skills} training for advanced technical roles. Personal profile variables and company size produced no relevant significant differences.
\textbf{Conclusion:} Personal profile variables do not determine the perception of quality and effectiveness, while professional trajectory variables (experience, level, area of work) produce localized differences. The voluntariness of participation remains a determining factor, in line with the literature. The absence of gender differences in a sample with 23\% women suggests that barriers operate before training, in access and representation, not during the learning experience.
\end{abstract}

\begin{CCSXML}
<ccs2012>
   <concept>
      <concept_id>10003456.10003457.10003521</concept_id>
      <concept_desc>Social and professional topics~Computing education</concept_desc>
      <concept_significance>500</concept_significance>
   </concept>
</ccs2012>
\end{CCSXML}

\ccsdesc[500]{Social and professional topics~Computing education}

\keywords{Corporate Training, Software Engineering, Sociodemographic Analysis, Gender, Quality Perception, Corporate Education}

\maketitle

% ============================================================
\section{Introduction}\label{sec:introducao}
% ============================================================

The software industry continuously invests in training and development programs, recognizing that a qualified and constantly evolving workforce constitutes a strategic advantage \cite{salas2012science, fitzgerald1992training}. Technology companies face growing pressure to build and maintain high-quality human capital, essential to sustain innovation and competitiveness in a sector characterized as training-intensive \cite{diniz2024skill}. However, despite these investments, many organizations lack mechanisms to assess the effectiveness of their programs \cite{santos2003employee}, and determining the return on investment in technology training remains an open problem \cite{coverstone2003training}.

This article is part of a multiphase research effort on corporate education in SE. In the first phase, a Systematic Mapping of the Literature identified significant gaps in research on training in SE \cite{de2025mapping}. In the second phase, a quantitative analysis with 282 Brazilian professionals identified the central factors of perceived quality and effectiveness of training \cite{siqueira2026quantitative}. In parallel, two qualitative analyses investigated, respectively, which learning experiences are perceived as most useful \cite{siqueira2026qualitative} and which improvements professionals suggest for corporate education processes \cite{siqueira2026professional}. All analyses were conducted on the same dataset, adopting complementary lenses.

These earlier findings identified \textit{what} matters in training. However, an equally relevant exploratory question remains open: does the professional's profile, that is, \textit{who} is trained, also influence perception? The literature on diversity and inclusion in SE documents systemic barriers faced by under-represented groups, such as women \cite{oliveira2024navigating}, and points out that individual factors such as experience and motivation can modulate learning outcomes \cite{salas2001science}. The literature on competencies in SE also highlights the need to understand how the essential competencies of professionals change over time \cite{assyne2022state}. In light of this, this phase investigates whether sociodemographic and professional variables produce significant differences in perception.

This study addresses the following research question:

\begin{quote}
    \textbf{RQ: To what extent do sociodemographic and professional variables influence the perception of quality and effectiveness of corporate training in Software Engineering?}
\end{quote}

In addition to answering the RQ, the analysis seeks to identify descriptive patterns in variables that eventually present significance, characterizing how and where the differences manifest themselves.

The main contributions of this study are:

\begin{enumerate}
    \item Empirical evidence on the influence of sociodemographic and professional variables, including gender, education, experience, and company size, on the perception of quality and effectiveness of corporate training in SE.
    \item The identification of a non-linear descriptive pattern in the relationship between length of experience and perception of impact, revealing a low-engagement zone between 3 and 6 years of experience.
    \item Evidence that the absence of gender differences, in a context of structural under-representation of women (23\% of the sample), suggests that gender barriers operate before training, not during.
\end{enumerate}

% ============================================================
\section{Theoretical Background}\label{sec:background}
% ============================================================

\subsection{The Salas \textit{Framework} and Antecedent Conditions}

This research is grounded in the \textit{The Science of Training} framework proposed by \citet{salas2001science} and revised by \citet{salas2012science}. The model organizes the training process into four dimensions: Needs Analysis, Antecedent Conditions, Instructional Methods, and Post-Training Conditions. For this phase, the most relevant dimension is \textbf{Antecedent Conditions}, which encompasses Individual Characteristics (cognitive ability, self-efficacy, goal orientation) and Training Motivation (shaped by individual and organizational variables). According to \citet{salas2001science}, these pre-training factors are as critical as the instructional method itself in determining outcomes.

The central question of this phase is precisely to test this premise: whether individual antecedent conditions, operationalized as sociodemographic variables, in fact produce measurable differences in perceived quality and effectiveness.

\subsection{Previous Phases of This Research}

The Systematic Mapping \cite{de2025mapping} analyzed 26 primary studies and revealed a concentration of literature on Instructional Methods and Strategies, with expressive gaps in subcategories such as \textit{Job/Task Analysis} (0 studies), \textit{Simulation-Based Training and Games} (0 studies), and \textit{Transfer of Training} (1 study). In the Antecedent Conditions dimension, \textit{Individual Characteristics} and \textit{Training Motivation} received limited attention (2 and 1 studies, respectively). This scarcity directly motivated this investigation.

The quantitative analysis \cite{siqueira2026quantitative}, with 282 Brazilian professionals, identified three factors strongly correlated with overall satisfaction (problem-solving reasoning, variety of activities, and instructor performance) and showed that mandatory participation reduces motivation and perception of quality.

Two complementary qualitative analyses were conducted on the same dataset. The first \cite{siqueira2026qualitative}, with 195 open-ended responses, identified that perceived usefulness is strongly linked to continuous technical updating and practical applicability. The second \cite{siqueira2026professional}, with 174 responses, revealed that practical applicability and pedagogical quality form an inseparable core of improvement.

In summary, the previous phases indicate that instructional design is the dominant factor in the perception of quality and effectiveness \cite{siqueira2026quantitative, siqueira2026qualitative, siqueira2026professional}. The next section examines what the literature presents about sociodemographic and professional factors, grounding the analysis conducted in this study.

\subsection{Sociodemographic and Professional Factors in the SE Training Literature}

The literature presents contrasting views on the influence of mandatoriness. \citet{gegenfurtner2016voluntary} argue that voluntariness favors autonomous motivation and learning transfer. In contrast, \citet{baldwin1991perils} suggest that mandatoriness signals strategic relevance, while excessive voluntariness may be interpreted as low institutional priority.

Regarding professional experience and career level, \citet{ohlmann2019perception} indicate that experienced professionals tend to evaluate decontextualized training as superfluous, and \citet{kupritz2002relative} demonstrate that resistance increases when content is perceived as irrelevant. \citet{assyne2022state} highlight the need for empirical research that understands how essential competencies change over the course of a career. This evidence suggests a complex relationship between length of experience and perceived value of training.

On gender in SE, \citet{oliveira2024navigating} document that Brazilian women face systemic challenges that persist from academia to industry, including insecurity, scarcity of female role models, stereotypes, and moral harassment. These barriers can influence not only access to training, but also how it is perceived.

In the field of competencies and area of work, the literature indicates that \textit{soft skills} are essential for software engineers at all levels and cannot be dissociated from the development process \cite{borges2024skills, galster2018toward}.

Regarding the relationship between academic background and professional preparation, \citet{borges2024skills} document that academic credentials and technical competencies are not sufficient for professional success, and that the gap between what is taught at the undergraduate level and what is demanded by industry persists as a central challenge. Formal education tends to privilege technical competencies over behavioral ones, although both are necessary for the practice of the profession. This gap motivates the investigation of whether educational level influences the perception of quality and effectiveness of corporate training.

% ============================================================
\section{Method}\label{sec:metodo}
% ============================================================

\subsection{Study Design}

This study constitutes the third phase of a sequential multiphase research effort on corporate education in SE \cite{de2025mapping, siqueira2026quantitative, siqueira2026qualitative, siqueira2026professional}. While the previous phases investigated \textit{what} determines the perception of quality and effectiveness, this phase investigates whether \textit{who} is trained also influences this perception. To this end, a significance analysis was conducted on the dataset collected in the previous phases, crossing 27 training perception items with 9 sociodemographic and professional variables, totaling 243 combinations.

\subsection{Instrument}

The instrument used in this study was developed and documented in detail in the quantitative phase of this research \cite{siqueira2026quantitative}. Construction followed an iterative four-stage process, guided by the dimensions of \citet{salas2001science}'s \textit{framework}: (1) extraction of 106 items from 20 sources in the literature; (2) semantic grouping into 40 items; (3) adaptation to a 5-point Likert scale into 30 items; and (4) cognitive load reduction to 27 items, validated in a pilot study with 10 SE professionals. The final instrument comprises three sections:

\textbf{Section A: Sociodemographic and professional variables (9 items).} Gender, age, state of residence, education level, length of experience in SE, professional level, company size, area of work, and nature of participation (mandatory or optional).

\textbf{Section B: Perception of quality and effectiveness (27 Likert items).} Items are organized into the four dimensions of the Salas framework: Needs Analysis (Q1--Q5), Antecedent Conditions (Q6--Q15), Instructional Methods and Strategies (Q16--Q20), and Post-Training Conditions (Q21--Q27). Each item is a declarative statement anchored in the participant's most recent learning experience, evaluated on a scale from 1 (\textit{Strongly disagree}) to 5 (\textit{Strongly agree}). The complete instrument is available in the artifact repository (Section~\ref{sec:artefatos}) and documented in \citet{siqueira2026quantitative}.

\textbf{Section C: Open-ended qualitative questions (2 optional items).} To complement the quantitative data and capture perceptions not contemplated by the closed items, the questionnaire included two open-ended free-response questions: (QQ1) ``What was the most useful learning experience you have ever had? Why?'' and (QQ2) ``If you could suggest an improvement to the learning processes, what would it be?''. Responding to these questions was optional. The responses to QQ1 and QQ2 were analyzed in dedicated complementary studies \cite{siqueira2026qualitative, siqueira2026professional} and are not the object of analysis in this article.

\subsection{Participants and Sampling}

The target population comprises professionals working in software engineering in Brazil, with higher education in Computing or related areas of Information and Communication Technology (ICT). Recruitment was by convenience, via professional social networks (LinkedIn) and messaging app groups (WhatsApp and Telegram). Participation was voluntary, without financial incentives.

Inclusion criteria:
\begin{itemize}
    \item being 18 years of age or older;
    \item working professionally in the area of software engineering;
    \item having participated, in the past 12 months, in at least one corporate training sponsored by the employing organization.
\end{itemize}

\subsection{Data Collection Procedure}

Collection took place entirely \textit{online}, from September 9 to November 2, 2025 ($\approx$8 weeks), through a questionnaire on the Google Forms platform. Completion of the consent, sociodemographic, and training characterization sections was mandatory. No data imputation technique was adopted. One response was excluded for not meeting the inclusion criteria, resulting in $n=282$ valid responses.

\subsection{Sociodemographic and Professional Profile of the Sample}

Regarding \textbf{gender}, 76.6\% ($n=216$) identified as male, 23.0\% ($n=65$) as female, and one participant (0.4\%) chose not to respond. The \textbf{age distribution} is unimodal and approximately symmetric ($\text{mean}=33.78$, $SD=7.73$, $\text{median}=33$ years).

Regarding \textbf{education level}, 49.7\% ($n=140$) hold a complete undergraduate degree and 38.7\% ($n=109$) hold a graduate degree; only 11.7\% ($n=33$) have not yet completed higher education.

Regarding \textbf{organization size}, 77.3\% ($n=218$) work in companies with 100 or more employees; medium-sized companies (50--99 employees) represent 11.7\% ($n=33$) and small companies (10--49 employees), 8.5\% ($n=24$). No professionals from microenterprises participated.

In terms of \textbf{area of work}, Development predominates (44.0\%, $n=124$), followed by People/Project Management (14.5\%, $n=41$), Quality Control (12.4\%, $n=35$), and Technical Leadership (8.5\%, $n=24$). Data Science (6.4\%), DevOps (6.0\%), and Architecture (3.9\%) are also represented.

Regarding \textbf{professional experience}, 43.6\% ($n=123$) have more than eight years in Software Engineering. The Senior level is the most frequent (28.7\%, $n=81$), followed by Mid-level (19.9\%, $n=56$) and Management/Leadership (19.5\%, $n=55$). Entry positions (Trainee and Junior) represent 16.4\% ($n=46$).

Regarding \textbf{geographic distribution}, the sample is concentrated in the Northeast (48.2\%, $n=136$) and Southeast (29.1\%, $n=82$), followed by the South (14.5\%, $n=41$), North (4.3\%, $n=12$), and Midwest (3.9\%, $n=11$).

Finally, the majority of participants (73\%, $n=206$) reported that participation in the most recent training was by personal choice, taking advantage of a benefit offered by the company.

The profile indicates a predominantly male sample (76.6\%), with high education (88.4\% with complete higher education or graduate studies), concentrated in large companies (77.3\%) and with strong representation from the Northeast (48.2\%) and Southeast (29.1\%). The mean age was 33.78 years ($SD=7.73$), with a median of 33 years.

\subsection{Data Analysis}

The analytical strategy consisted of systematically testing whether each of the 9 sociodemographic variables produces significant differences in the 27 perception items ($9 \times 27 = 243$ combinations). Given the ordinal nature of the data (5-point Likert scale) and the absence of guaranteed normality in the distributions, non-parametric tests were chosen \cite{siegel2006estatistica}.

The choice of statistical test was determined by the nature of each sociodemographic variable:

\begin{itemize}
    \item \textbf{Kruskal-Wallis test \cite{kruskal1952use}:} applied for all categorical variables: gender, nature of participation (mandatory \textit{vs.} optional), education level, length of experience, professional level, company size, area of work, and state/region. For variables with two groups (gender and mandatoriness), the Kruskal-Wallis test is mathematically equivalent to the Mann-Whitney test \cite{mann1947test}. Groups with $n < 5$ were excluded from comparisons to ensure robustness.
    \item \textbf{Spearman correlation \cite{spearman1961proof}:} applied for the relationship between age and each perception item. Unlike the other sociodemographic variables, age presents an approximately normal distribution ($\text{mean}=33.78$, $SD=7.73$, $\text{median}=33$) and was treated as a continuous variable, without a priori categorization, avoiding loss of information from grouping.
\end{itemize}

The significance level adopted was $\alpha = 0.05$, a convention widely accepted in social and behavioral sciences research. For Kruskal-Wallis analyses with significant results ($p < 0.05$), pairwise \textit{post-hoc} comparisons with Dunn's test \cite{dunn1964multiple} were conducted to identify which specific groups differ from each other. Bonferroni correction \cite{bland1995multiple} was applied to the \textit{post-hoc} p-values, adjusting the significance threshold by the number of comparisons performed and controlling the risk of false positives.

Effect size, which indicates the practical magnitude of the difference regardless of statistical significance, was calculated using Epsilon-squared ($\varepsilon^2 = H / (n-1)$) for Kruskal-Wallis and the coefficient $r$ itself for Spearman, classified as: small ($< 0.06$), medium ($0.06$ to $0.14$), and large ($\geq 0.14$).

For interpretation purposes, the 9 sociodemographic variables were organized into three analytical groups: (1) \textbf{personal profile}, comprising intrinsic characteristics of the individual (gender, age, and education level); (2) \textbf{professional trajectory}, comprising variables linked to the moment in the career (length of experience, professional level, and area of work); and (3) \textbf{organizational context}, comprising characteristics of the company and the training (company size, state/region, and nature of participation). This categorization allows distinguishing whether the perception of training is influenced by who the professional is, where they are in their career, or the conditions in which the training is offered.

It is important to note that, although the Bonferroni correction was applied in the \textit{post-hoc} tests, the 243 top-level tests use $\alpha=0.05$ without global correction, reflecting the exploratory nature of the study. A detailed analysis of the impact of this choice is presented in Section~\ref{sec:resultados}.

\subsection{Ethical Considerations}

The study was approved by the Research Ethics Committee (CAEE: 91121125.4.0000.5208; Opinion No. 7,816,810). Participation was voluntary, and data were collected anonymously and confidentially. Participants could withdraw at any time without prejudice. The complete questionnaire, analysis \textit{scripts}, and anonymized data are available in the artifact repository (Section~\ref{sec:artefatos}).

% ============================================================
\section{Results}\label{sec:resultados}
% ============================================================

\subsection{Overview: The Significance Matrix}

\begin{figure}[htbp]
    \centering
    \includegraphics[width=1\linewidth]{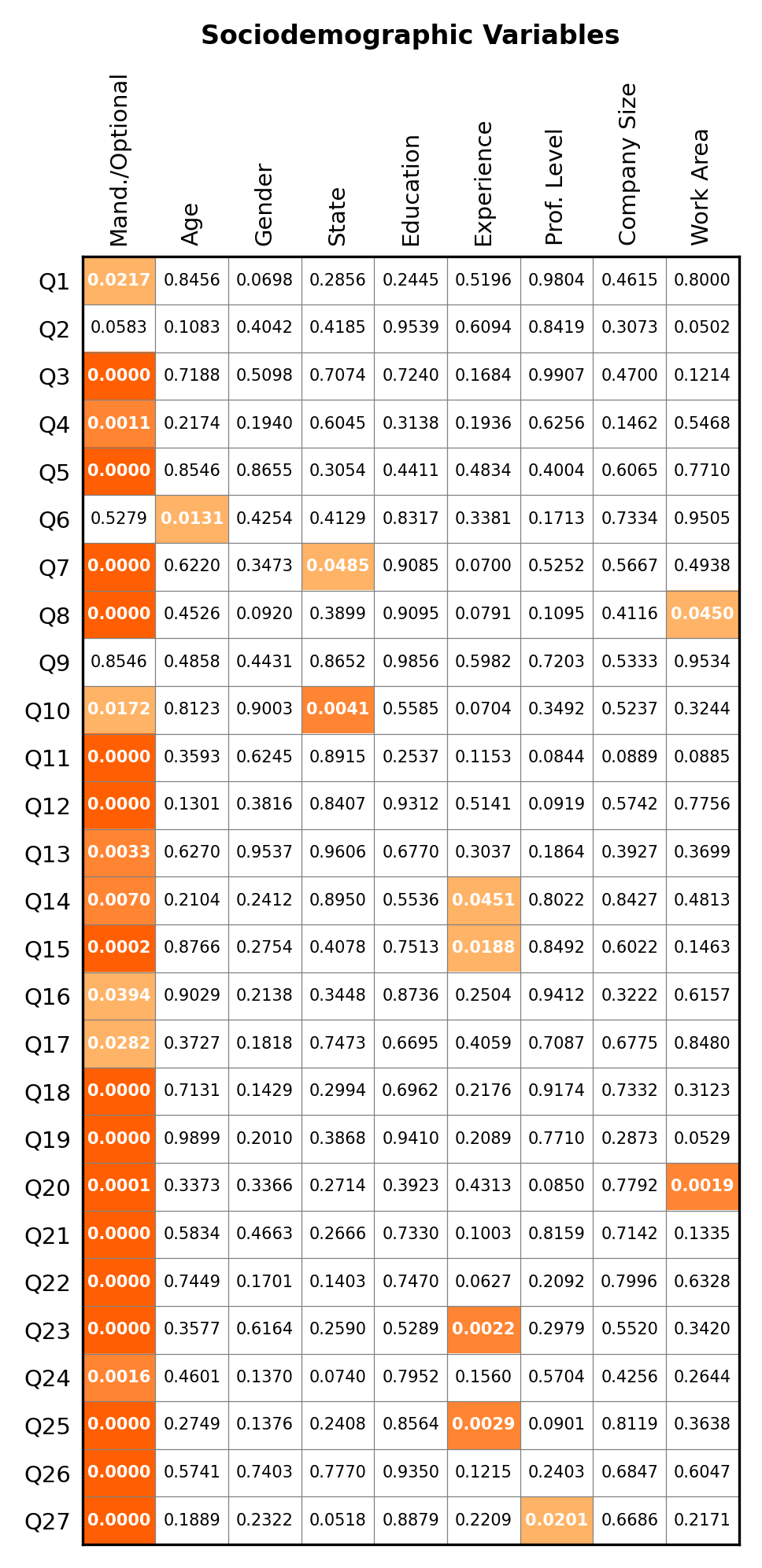}
    \caption{Significance matrix.\\
    {\scriptsize (1) Columns: sociodemographic variables; rows: perception items. (2) Orange indicates $p < 0.05$.}}
    \label{fig:significancia}
\end{figure}

Figure~\ref{fig:significancia} presents the significance matrix resulting from the crossing of 27 perception items with 9 sociodemographic variables, totaling 243 combinations tested. Color intensity indicates the level of statistical significance.

A methodological caveat is necessary before interpreting the results. With 243 tests conducted at $\alpha=0.05$, approximately 12 significant results would be expected by chance ($243 \times 0.05 \approx 12$). Of the 243 combinations, 35 showed statistical significance. Of this total, 24 are concentrated in the variable nature of participation (mandatory \textit{vs.} optional), whose effect is widely documented in the literature \cite{gegenfurtner2016voluntary} and in the previous phases of this research \cite{siqueira2026quantitative}. No result was excluded from the analysis: all 35 are reported in full in the following subsections. However, when separating the 24 results from mandatoriness, 11 significant results remain among the other 216 combinations, a number compatible with what is expected by chance ($216 \times 0.05 \approx 11$).

The most expressive result is the concentration of significance in the \textbf{Mandatory/Optional} column, with 24 of 27 items showing $p < 0.05$. In contrast, the other columns are predominantly white. Only 11 of the remaining 216 combinations reached significance. The \textbf{personal profile} variables and \textbf{company size} did not produce relevant significant differences, with the exception of a small-effect correlation between age and Q6 (Impact on personal time).

For the variables that showed significance, pairwise \textit{post-hoc} comparisons (Dunn with Bonferroni correction) were conducted to identify which groups differ from each other, and the effect size ($\varepsilon^2$) was calculated for each significant result. The following subsections report the global results, the pairs identified by the \textit{post-hoc}, and the corresponding effect sizes.

\subsection{Personal Profile and Company Size: Absence of Significance}\label{sec:genero}

The gender column in Figure~\ref{fig:significancia} is entirely white: none of the 27 items showed a significant difference between men (76.6\%, $n=216$) and women (23.0\%, $n=65$).

Likewise, the variables \textbf{education level} and \textbf{company size} did not produce significant differences in any item. The sample is concentrated on professionals with complete higher education or graduate studies (88.4\%) and in companies with 100+ employees (77.3\%).

\subsection{Length of Experience and Perception of Impact: A Non-Linear Pattern}\label{sec:experiencia}

\begin{figure}[htbp]
    \centering
    \includegraphics[width=0.8\linewidth]{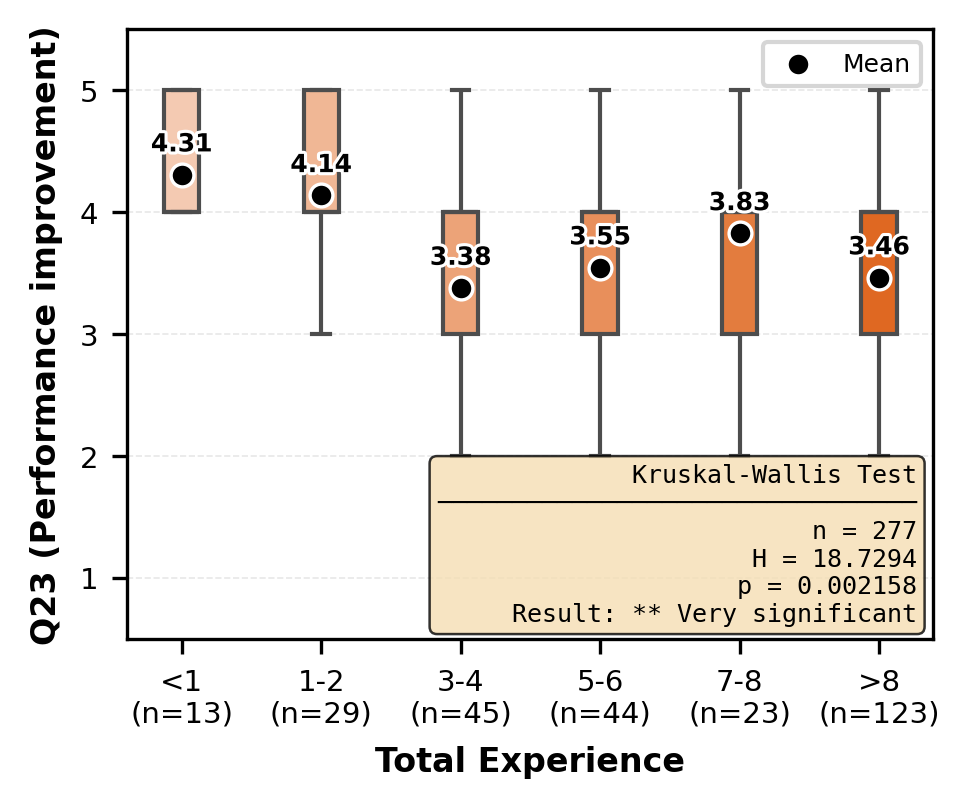}
    \caption{Q23 (Performance improvement) by Length of Experience.\\
    {\scriptsize Kruskal-Wallis: $H=18.73$, $p=0.002$.}}
    \label{fig:Q23}
\end{figure}

\begin{figure}[htbp]
    \centering
    \includegraphics[width=0.8\linewidth]{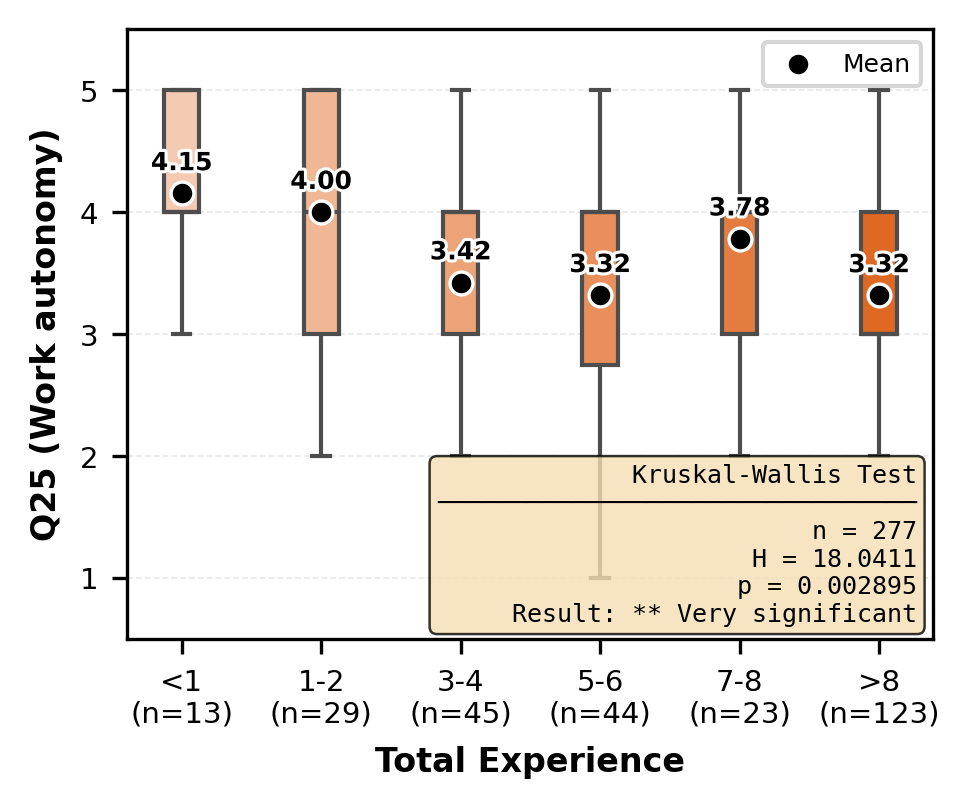}
    \caption{Q25 (Autonomy at work) by Length of Experience.\\
    {\scriptsize Kruskal-Wallis: $H=18.04$, $p=0.003$.}}
    \label{fig:Q25}
\end{figure}

Length of experience in software engineering stood out as the second most influential factor. Figures~\ref{fig:Q23} and~\ref{fig:Q25} illustrate the two items with the largest effect, and Figure~\ref{fig:significancia} points to significant results for items Q14 (Organization and structure, $p=0.045$, $\varepsilon^2=0.041$, small effect), Q15 (Useful materials, $p=0.019$, $\varepsilon^2=0.049$, small effect), Q23 (Performance improvement, $p=0.002$, $\varepsilon^2=0.068$, medium effect), and Q25 (Autonomy at work, $p=0.003$, $\varepsilon^2=0.065$, medium effect).

\begin{figure}[htbp]
    \centering
    \includegraphics[width=0.8\linewidth]{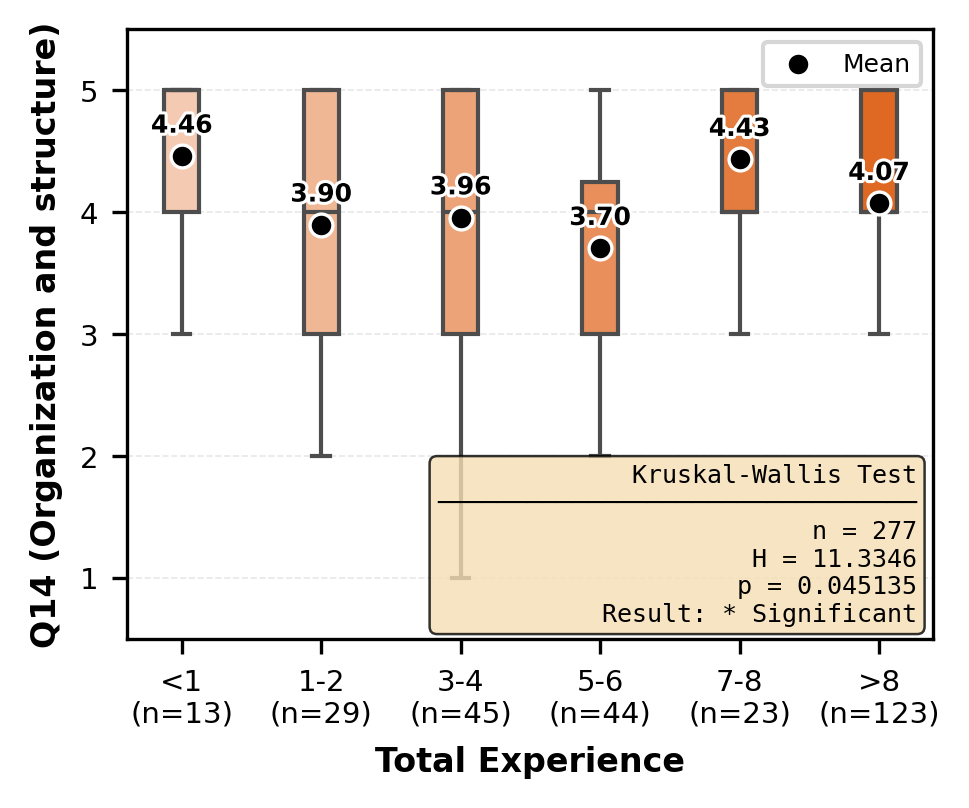}
    \caption{Q14 (Organization and structure) by Length of Experience.}
    \label{fig:Q14}
\end{figure}

\begin{figure}[htbp]
    \centering
    \includegraphics[width=0.8\linewidth]{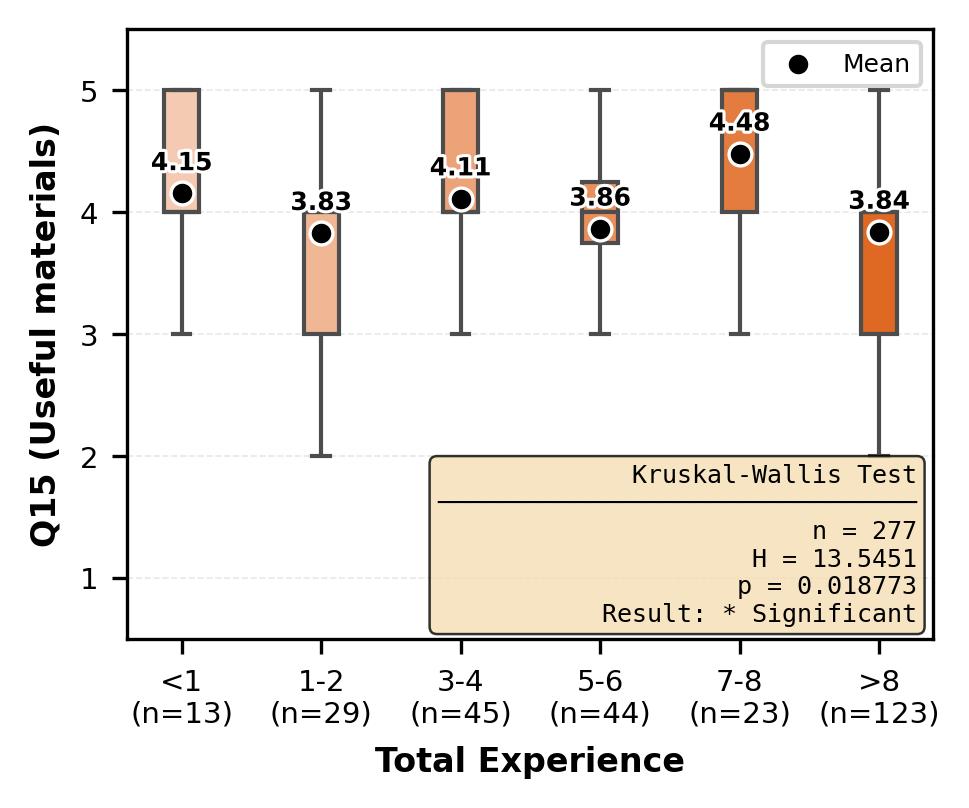}
    \caption{Q15 (Useful materials) by Length of Experience.}
    \label{fig:Q15}
\end{figure}

The analysis of the four significant items (Figures~\ref{fig:Q23},~\ref{fig:Q25},~\ref{fig:Q14}, and~\ref{fig:Q15}) reveals a consistent non-linear descriptive pattern: the highest means are concentrated in the up-to-2-years experience bands, followed by a drop in the 3-to-6-year bands, a recovery in the 7--8 years group ($n=23$), and a new drop in professionals with more than 8 years (44.4\%, $n=123$).

The \textit{post-hoc} comparisons (Dunn with Bonferroni correction) identified the following significant pairs. For Q23 (Performance improvement): up to 1 year \textit{vs.} 3--4 years ($p=0.050$), 1--2 years \textit{vs.} 3--4 years ($p=0.042$), and 1--2 years \textit{vs.} more than 8 years ($p=0.031$). For Q25 (Autonomy at work): 1--2 years \textit{vs.} more than 8 years ($p=0.024$). For Q15 (Useful materials): 7--8 years \textit{vs.} more than 8 years ($p=0.022$). For Q14 (Organization and structure), no pair survived the Bonferroni correction.

\subsection{Age and Personal Time: A Generational Sensitivity}

\begin{figure}[htbp]
    \centering
    \includegraphics[width=0.8\linewidth]{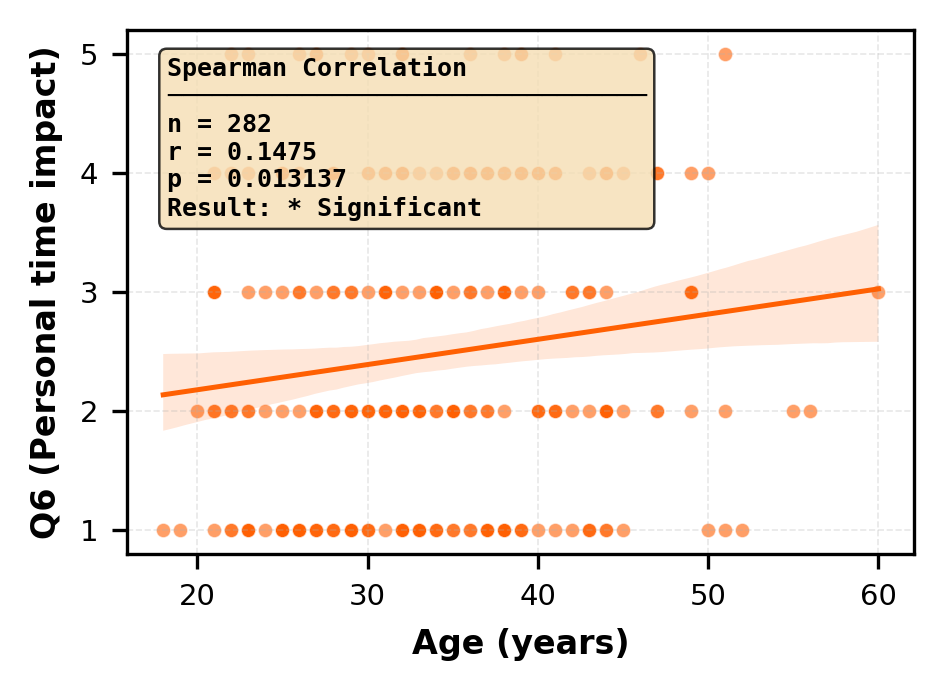}
    \caption{Q6 (Impact on personal time) by Age.\\
    {\scriptsize Spearman: $r=0.147$, $p=0.013$.}}
    \label{fig:Q6}
\end{figure}

The analysis of the relationship between age and item Q6 (Impact on personal time) revealed a significant positive correlation (Spearman $r=0.147$, $p=0.013$), as illustrated in Figure~\ref{fig:Q6}.

\begin{figure}[htbp]
    \centering
    \includegraphics[width=0.8\linewidth]{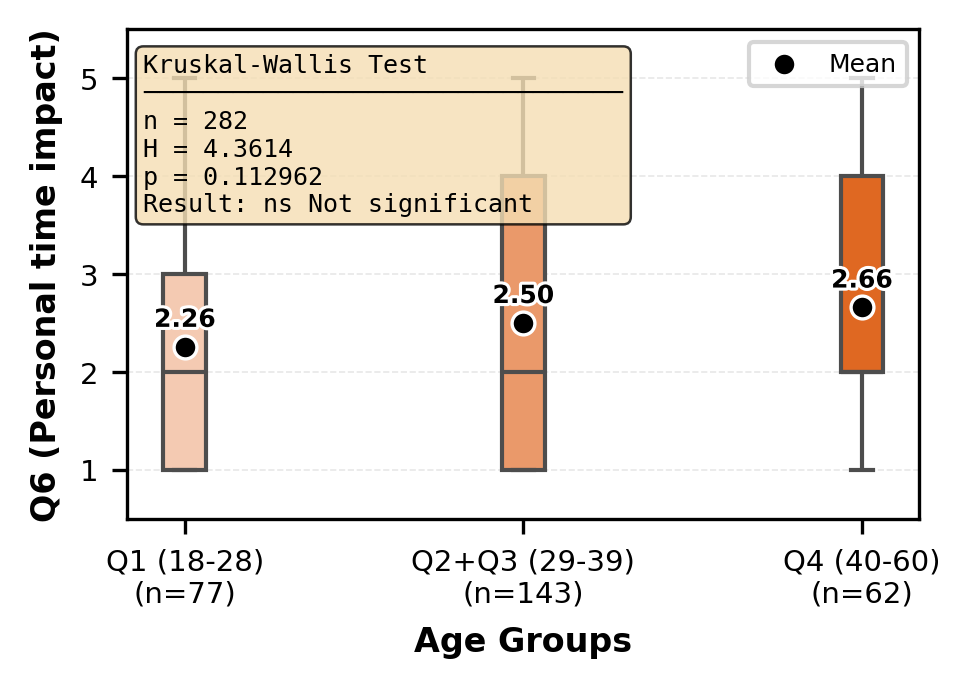}
    \caption{Q6 (Impact on personal time) by age groups.}
    \label{fig:Q6_agrupado}
\end{figure}

When segmented by age groups (Figure~\ref{fig:Q6_agrupado}), the descriptive trend holds: young people up to 28 years old presented the lowest mean ($2.26$), while professionals over 40 recorded the highest ($2.66$), although the difference between groups does not reach significance ($p=0.113$). The effect size is small ($r=0.147$).

\subsection{Regional Disparities: Conditions of Access, Not Instructional Quality}

\begin{figure}[htbp]
    \centering
    \includegraphics[width=0.8\linewidth]{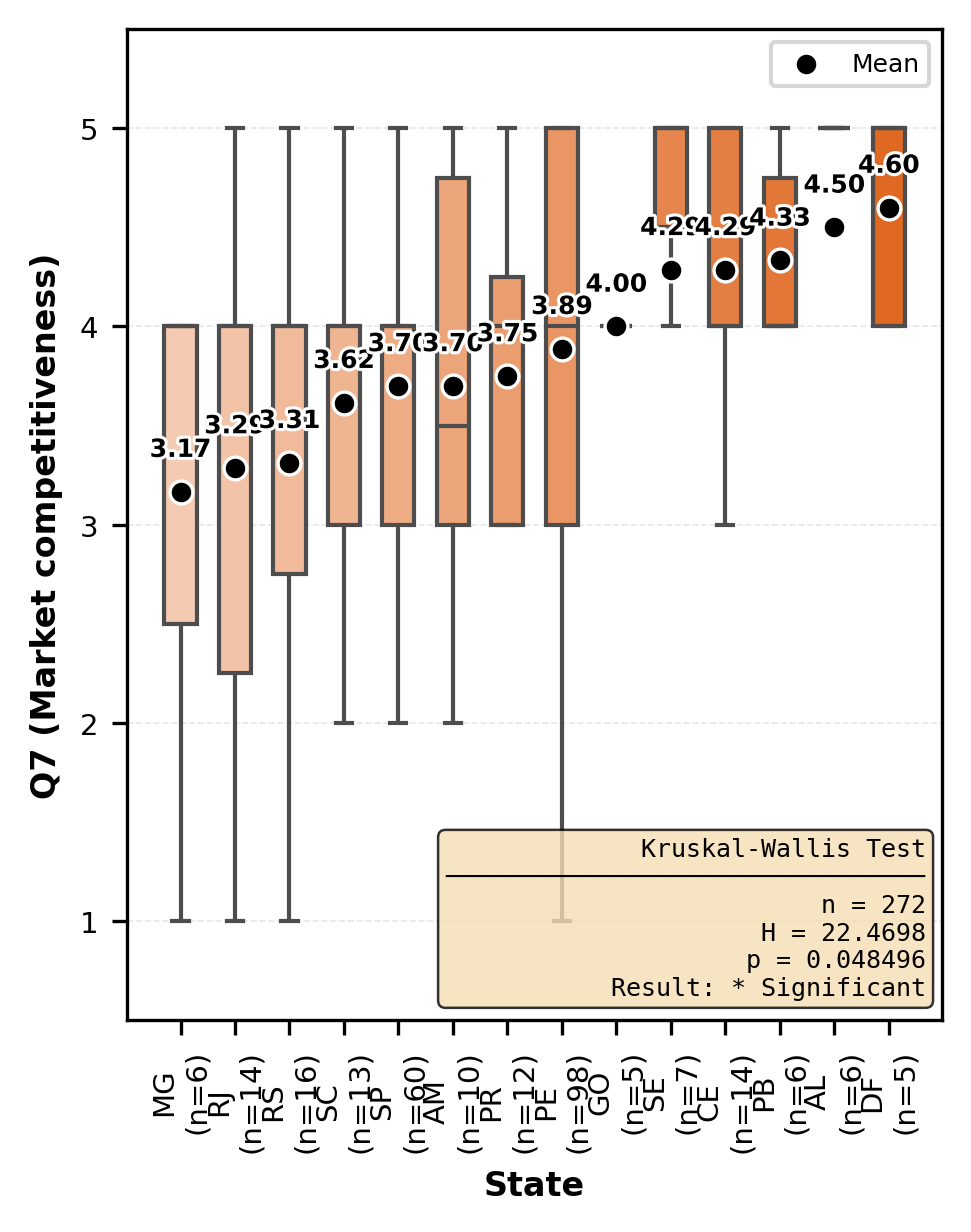}
    \caption{Q7 (Relevance for competitiveness) by State.}
    \label{fig:Q7_estado}
\end{figure}

\begin{figure}[htbp]
    \centering
    \includegraphics[width=0.8\linewidth]{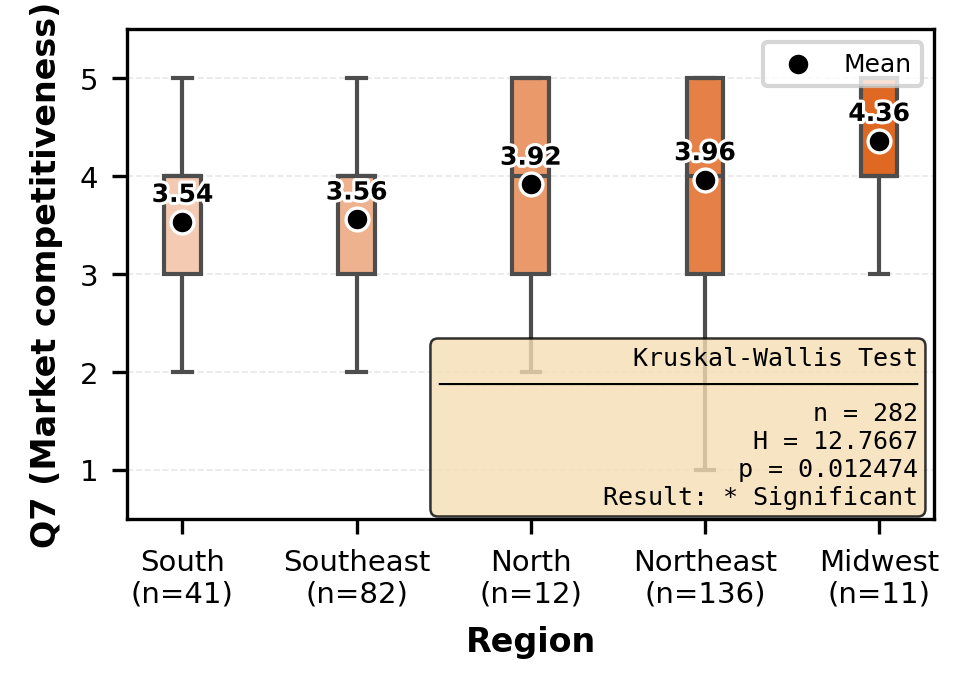}
    \caption{Q7 (Relevance for competitiveness) by Region.\\
    {\scriptsize Kruskal-Wallis: $p=0.012$.}}
    \label{fig:Q7_regiao}
\end{figure}

The analysis by state revealed significant differences in two items (Figures~\ref{fig:Q7_estado} and~\ref{fig:Q7_regiao}): Q7 (Relevance for competitiveness, $p=0.048$, $\varepsilon^2=0.083$, medium effect) and Q10 (Training during working hours, $p=0.004$, $\varepsilon^2=0.112$, medium effect).

\begin{figure}[htbp]
    \centering
    \includegraphics[width=0.8\linewidth]{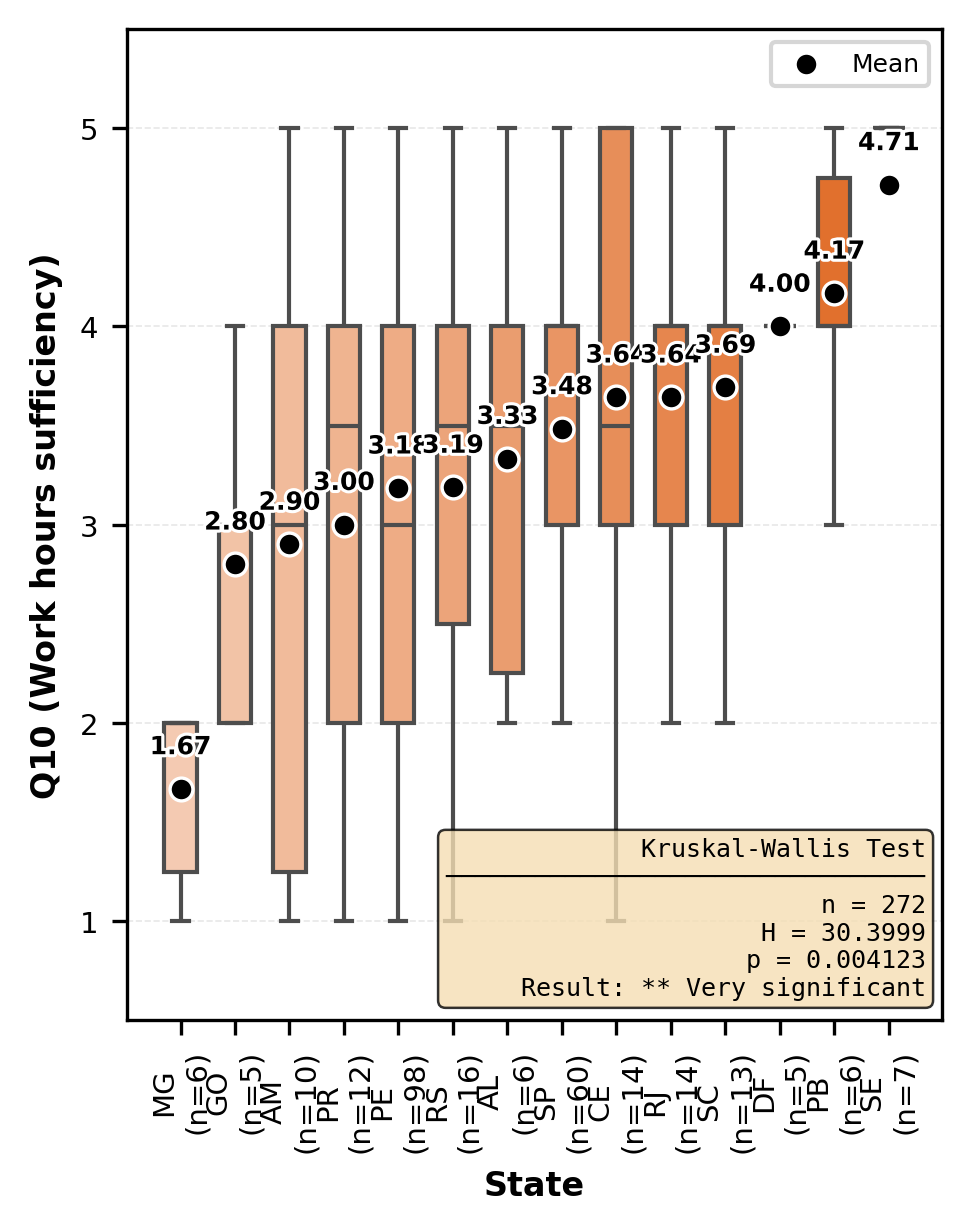}
    \caption{Q10 (Training during working hours) by State.\\
    {\scriptsize Kruskal-Wallis: $p=0.004$.}}
    \label{fig:Q10_estado}
\end{figure}

When grouped by macro-region, the difference in Q7 (Relevance for competitiveness) remained significant ($p=0.012$): South ($3.54$) and Southeast ($3.56$) presented the lowest indicators, while the Midwest obtained the highest mean ($4.36$). For Q10 (Training during working hours), whose distribution by state is illustrated in Figure~\ref{fig:Q10_estado}, the significance diluted at the regional level ($p=0.563$), although the Midwest maintained the highest nominal mean.

The \textit{post-hoc} comparison for Q10 (Training during working hours) identified the pair Minas Gerais \textit{vs.} Sergipe ($p=0.001$) as the only difference that survived the Bonferroni correction. For Q7 (Relevance for competitiveness), no isolated pair reached significance in the \textit{post-hoc}. It is important to note that the state subgroups have small sample sizes (MG: $n=6$; DF: $n=5$; SE: $n=7$), which limits the robustness of these comparisons.

\subsection{Area of Work: The \textit{Soft Skills} Gap in Technical Roles}

\begin{figure}[htbp]
    \centering
    \includegraphics[width=0.8\linewidth]{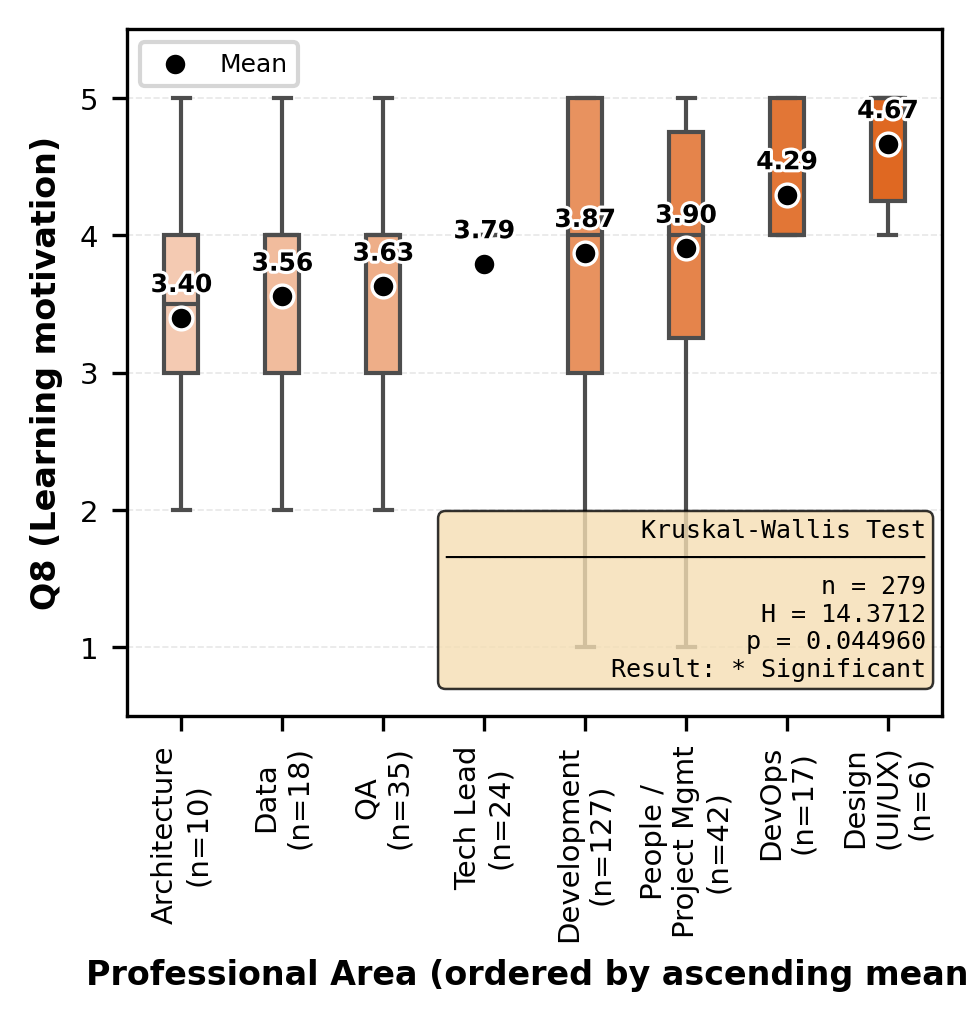}
    \caption{Q8 (Motivation to learn) by Area of Work.\\
    {\scriptsize Kruskal-Wallis: $p=0.045$.}}
    \label{fig:Q8}
\end{figure}

\begin{figure}[htbp]
    \centering
    \includegraphics[width=0.8\linewidth]{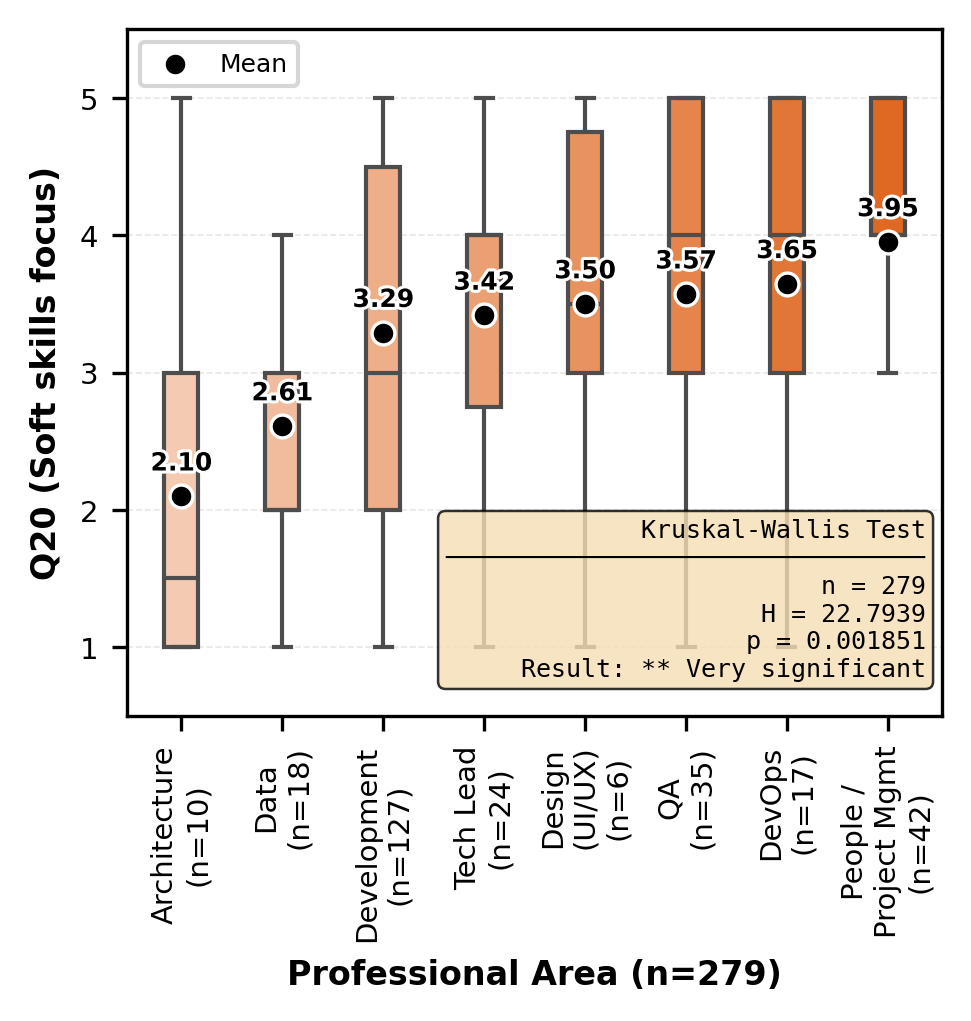}
    \caption{Q20 (Soft skills focus) by Area of Work.\\
    {\scriptsize Kruskal-Wallis: $p=0.002$.}}
    \label{fig:Q20}
\end{figure}

Area of work produced significant differences in two items (Figures~\ref{fig:Q8} and~\ref{fig:Q20}): Q8 (Motivation to learn, $p=0.045$, $\varepsilon^2=0.052$, small effect) and Q20 (Soft skills focus, $p=0.002$, $\varepsilon^2=0.082$, medium effect).

The Design (UI/UX) ($4.67$) and DevOps ($4.29$) areas presented the highest motivation indices, while Software Architecture obtained the lowest mean ($3.40$).

The Q20 (Soft skills focus) scenario, detailed in Figure~\ref{fig:Q20}, presents expressive variation across areas. Architecture recorded the lowest mean ($2.10$), while People/Project Management obtained the highest ($3.95$). Development (45.0\%, $n=127$) showed an intermediate mean ($3.29$) with the largest dispersion among the groups.

The \textit{post-hoc} comparisons for Q20 (Soft skills focus) identified two significant pairs: Architecture \textit{vs.} People/Project Management ($p=0.006$) and People/Project Management \textit{vs.} Data ($p=0.012$). For Q8 (Motivation to learn), no pair survived the Bonferroni correction.

\subsection{Mandatory \textit{versus} Optional: Confirming Previous Findings}\label{sec:obrigatoriedade}

Training mandatoriness was the most influential factor, with 24 of 27 items showing significant differences. Effect size ranged from small to large, with a highlight for Q11 (Obligation over interest, $\varepsilon^2=0.240$, large effect) and Q21 (Overall satisfaction, $\varepsilon^2=0.145$, large effect). Table~\ref{tab:obrig_top5} presents the five items with the largest disparities.

\begin{table}[htbp]
\centering
\caption{Five largest disparities between mandatory and optional participation}
\label{tab:obrig_top5}
\small
\begin{tabular}{llccc}
\toprule
\textbf{Item} & \textbf{Description} & \textbf{Mand.} & \textbf{Opt.} & \textbf{$\Delta$} \\
\midrule
Q22 & Problem-solving reasoning & 2.97 & 3.94 & 0.97 \\
Q7 & Relevance for competitiveness & 3.09 & 4.06 & 0.97 \\
Q8 & Motivation to learn & 3.17 & 4.10 & 0.93 \\
Q5 & Consideration of employee opinion & 2.58 & 3.48 & 0.90 \\
Q21 & Overall satisfaction & 3.25 & 4.14 & 0.89 \\
\bottomrule
\end{tabular}
\end{table}

The three non-significant exceptions were Q2 (Support and resources, $p=0.058$), Q6 (Impact on personal time, $p=0.528$), and Q9 (Leadership encouragement, $p=0.855$).

\subsection{Professional Level: Growth Perceived by Juniors}

The analysis by professional level revealed a significant difference only in Q27 (Career growth opportunities, $p=0.020$, $\varepsilon^2=0.048$, small effect).

\begin{figure}[htbp]
    \centering
    \includegraphics[width=0.8\linewidth]{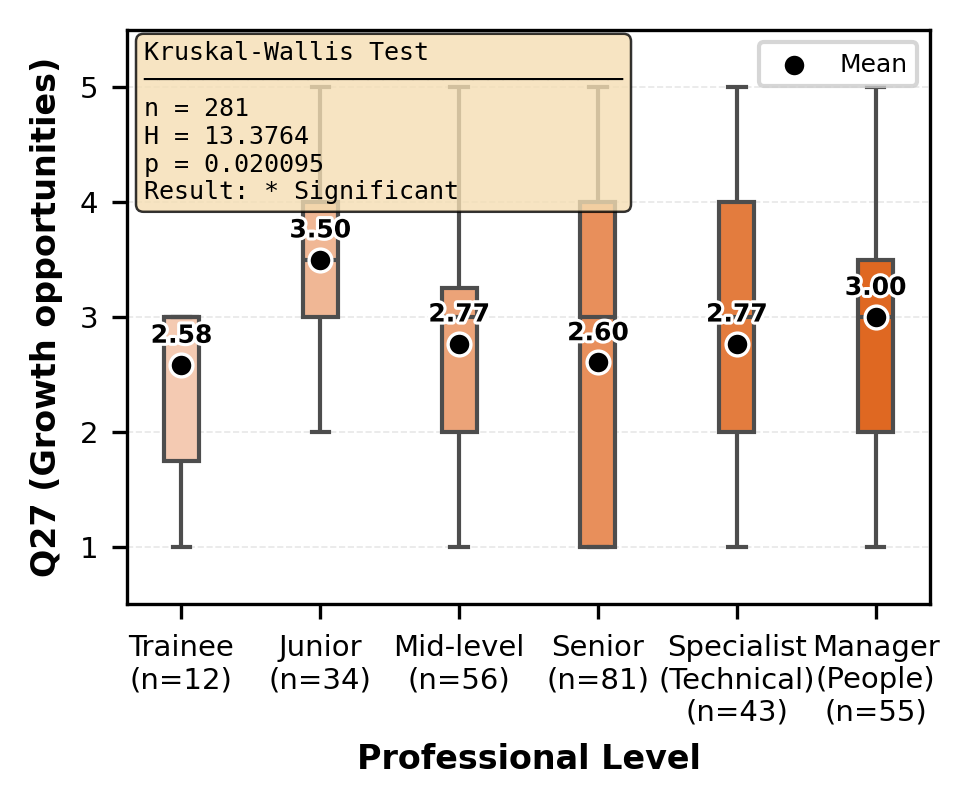}
    \caption{Q27 (Career growth opportunities) by Professional Level.\\
    {\scriptsize Kruskal-Wallis: $p=0.020$.}}
    \label{fig:Q27}
\end{figure}

As illustrated in Figure~\ref{fig:Q27}, the Junior level showed the highest agreement ($3.50$), while trainee respondents recorded the lowest value ($2.58$). For Mid-level, Senior, and Specialist professionals, the means remained close to neutrality. The \textit{post-hoc} comparison identified the pair Junior \textit{vs.} Senior ($p=0.009$) as the only difference that survived the Bonferroni correction.

% ============================================================
\section{Discussion}\label{sec:discussao}
% ============================================================

\subsection{It's Not About Who, But When in the Career}

\textit{In response to the \textbf{RQ}, the data indicate that personal profile variables do not influence the perception of quality and effectiveness of training, while professional trajectory variables produce localized differences. It is not about who the professional is, but at which point in their career they find themselves.}

The central finding of this study is that the perception of quality and effectiveness of corporate training in SE is not determined by the personal profile of the participant. Personal profile variables produced no relevant significant differences, just as company size did not. This is counterintuitive in light of the diversity and inclusion literature, but consistent with the findings of the previous phases of this research \cite{siqueira2026quantitative, siqueira2026qualitative, siqueira2026professional}.

In dialogue with \citet{salas2001science}'s \textit{framework}, the results suggest that individual Antecedent Conditions of personal profile do not influence perception, while variables linked to professional trajectory produce localized differences. The combined evidence supports the thesis of the title: it is not about who the professional is, but at which point in their career they find themselves. This finding may help explain the uneven distribution found in the systematic mapping by \citet{de2025mapping}, where the \textit{Specific Learning Approaches} subcategory concentrated the largest number of studies (6), while \textit{Individual Characteristics} (2 studies) and \textit{Training Motivation} (1 study) received limited attention.

\subsection{Gender: Barriers Before, Not During}

The absence of gender differences is the most provocative finding. With 76.6\% of the sample being male, it would be reasonable to expect the 23\% of women to report distinct experiences, given the systemic barriers documented by \citet{oliveira2024navigating}. The perceptual homogeneity suggests that selection occurs before training: women who reached the point of participating in corporate training in SE have already overcome entry barriers in the sector. Once inside, their perceptions converge.

This does not mean that gender barriers do not exist in SE training. It means they operate before, in under-representation, in access, and in the conditions that allow reaching training, and not during the learning experience. This interpretation should be framed as a hypothesis: to confirm that perceptual homogeneity results from pre-training filtering, it would be necessary to compare with data on professionals who did not reach training, which this study does not have. In addition, the low female representation in the sample (23\%, $n=65$) constitutes simultaneously evidence of the social problem of under-representation and a methodological limitation, since statistical power to detect small effects between groups may be insufficient.

The same logic applies to the absence of significance in education level and company size. The concentration of the sample in professionals with higher education (88.4\%) and in large companies (77.3\%) compresses variance in these groups, which may explain the absence of differences. \textit{Brazilian corporate training in SE seems to operate as a homogenized experience for those who already have access to it.}

\subsection{Segment by Stage, Not by Profile}

Professional experience was the only variable with consistent impact on multiple items, and the non-linear pattern is the most original finding of this study. The cross analysis between length of experience and professional level in the sample indicates that professional level increases consistently with length of work, with progressive growth between entry positions and the mid-level, followed by greater variability among senior professionals. \textit{In this context, the 3-to-6-year band corresponds predominantly to the mid-level, configuring a ``dead zone'' where generic training seems not to add perceptible value:} the professional has already overcome the initial learning curve, but the content offered does not yet reflect the more sophisticated demands of the next stage. This pattern converges with \citet{ohlmann2019perception}, who indicate that experienced professionals perceive generic training as superfluous, and with \citet{kupritz2002relative}, who demonstrate that resistance increases in the face of irrelevant content.

\textit{The recovery at 7--8 years ($n=23$) may reflect the transition to senior or leadership positions, where new competencies are demanded.} The ceiling effect in professionals with more than 8 years suggests that standard training exhausts its potential to add value to the specialist. This finding empirically addresses the gap identified by \citet{assyne2022state}, who called precisely for a better understanding of how essential competencies change over the course of a career. The same pattern is observed in the analysis by professional level: the Junior \textit{vs.} Senior pair was the only significant difference in Q27 (Career growth opportunities), \textit{indicating that the perception of training as a career lever decreases as the professional advances in the hierarchy.}

\textit{The practical implication is clear: training programs should be segmented not by personal profile (gender, age, education level), but by career stage. Personalization should target the professional moment, not personal characteristics.}

It is necessary, however, to acknowledge a confounding variable that may influence the interpretation of this pattern. The questionnaire does not differentiate types of corporate training (e.g., \textit{onboarding}, technical updating, \textit{compliance}), and different experience profiles may be evaluating distinct types of capacitation. For example, professionals at the start of their careers may be evaluating predominantly \textit{onboarding} experiences, whose introductory nature tends to produce high perceptions of gain. In this scenario, the observed pattern would reflect, at least partially, differences in the type of training received by each band. Although the descriptive pattern is consistent across the four significant items and the \textit{post-hoc} tests identify specific pairs, future studies should include typological controls to isolate the effect of experience from the effect of training type.

\subsection{Structural Conditions and Content Gaps}

\textit{Regional differences reflect access conditions (protected time, perception of competitiveness), not instructional quality.} The two items significant by state (Q7 and Q10) refer to structural conditions, not pedagogical quality dimensions. This converges with the analysis of improvement suggestions \cite{siqueira2026professional}, which identified ``Time and structural conditions'' as one of the five priority dimensions.

\textit{The positive correlation between age and perception of damage to personal time (Q6), although with a small effect, suggests that older professionals, possibly with greater family responsibilities, tend to be more sensitive to training's invasion of their off-hours.}

Regarding mandatoriness, the three non-significant exceptions (Q2, Q6, and Q9) are revealing. \textit{These exceptions suggest that, although mandatoriness impacts the perception of training, it does not alter the view of structural conditions. Participants recognize the support given by the company regardless of the motivation for their participation.} These results are consistent with the literature \cite{gegenfurtner2016voluntary} and with previous findings of this research \cite{siqueira2026quantitative, siqueira2026professional}.

The \textit{soft skills} gap in advanced technical roles, especially in Architecture, is a finding with direct implications for program design. \citet{borges2024skills} identified that \textit{soft skills} such as communication, teamwork, and leadership are essential for software engineers at all levels. \textit{The low motivation in Architecture suggests that the content offered may be excessively basic or theoretical for the seniority level of these professionals.} The high dispersion in Development, in turn, \textit{suggests an absence of standardization: while some developers receive strong behavioral training, others experience purely technical content.} The data suggest that Brazilian industry concentrates behavioral training in management roles, neglecting technical professionals who also need to ``sell'' decisions and negotiate with different teams.

% ============================================================
\section{Threats to Validity and Limitations}\label{sec:limitacoes}
% ============================================================

\textbf{External validity:} The sample is restricted to Brazilian professionals recruited by convenience, with predominance from the Northeast (48.2\%) and Southeast (29.1\%). Generalization to other cultural or economic contexts is limited.

\textbf{Selection bias:} The male predominance (76.6\%) and high education (88.4\% with higher education) suggest that respondents represent professionals who have already overcome entry barriers in the sector, which may homogenize the perceptions reported. The binary categorization of gender (male/female/not informed) is a limitation: an ``other'' option or open field would allow capturing diverse gender identities.

\textbf{Sample sizes:} Subgroups such as states (MG: $n=6$; SE: $n=7$) and areas of work (Design: $n=6$; Architecture: $n=10$) have limited sizes, which reduces statistical power and the reliability of the estimates. Findings involving these subgroups should be interpreted as exploratory.

\textbf{Multiple comparisons:} With 243 combinations tested, there is a risk of false positives. Although Bonferroni correction was applied in the \textit{post-hoc} tests, top-level tests use $\alpha=0.05$ without global correction, which should be considered in interpretation.

\textbf{Shared dataset and intra-study triangulation:} This article analyzes the same dataset ($n=282$) used in previous analyses: exploratory \cite{siqueira2026quantitative}, qualitative on useful experiences \cite{siqueira2026qualitative}, and qualitative on improvement suggestions \cite{siqueira2026professional}. The convergence between the sociodemographic findings (this study), exploratory, and qualitative findings does not constitute triangulation between independent samples, but intra-study methodological triangulation, in which each article adopts a distinct analytical lens on the same data. Although the consistency between results strengthens the internal validity of the findings, the sharing of the dataset limits the independence of the evidence and prevents claims of replication.

\textbf{Memory and self-report bias:} Participants reported the most recent training experience, subject to recency or forgetting distortions. Professional level categories are self-declared and may be interpreted heterogeneously.

\textbf{Construct validity with single items:} Each construct (e.g., perceived quality, instructor performance, overall satisfaction) was assessed by a single Likert item, which may not fully capture its multidimensional nature. Although single-item measures are common in large-scale \textit{surveys}, they limit the assessment of internal consistency. The absence of significance in some sociodemographic variables may, therefore, reflect measurement limitations and not real homogeneity. Future studies should adopt multi-item scales and confirmatory factor analysis.

\textbf{Absence of differentiation by training type:} As discussed in Section~\ref{sec:discussao}, the questionnaire does not distinguish between different types of corporate training, which constitutes a confounding variable that limits the interpretation of findings on experience. Future studies should include typological controls or analyses stratified by training type.

\textbf{Absence of objective metrics:} This study is based entirely on self-reported perceptions. In the context of sociodemographic analyses, this limitation is particularly relevant because different groups may present distinct response patterns (e.g., cultural tendency toward agreement) regardless of actual training experience. Future work should complement \textit{survey} data with objective effectiveness metrics, such as productivity or code quality indicators.

% ============================================================
\section{Conclusion}\label{sec:conclusao}
% ============================================================

This study investigated whether sociodemographic and professional variables influence the perception of quality and effectiveness of corporate training in Software Engineering, crossing 27 perception items with 9 sociodemographic variables on a sample of 282 Brazilian professionals.

In response to the \textbf{RQ}, personal profile variables and company size produced no relevant significant differences. Training mandatoriness was the dominant factor, confirming previous findings \cite{siqueira2026quantitative, siqueira2026professional}. Professional trajectory variables (experience, area of work, and professional level) and regional conditions produced localized differences. The most relevant pattern was a non-linear descriptive pattern in the relationship between experience and perception of impact, with a low-engagement zone between 3 and 6 years and a transient recovery at 7--8 years. Regional differences reflect structural conditions of access. Differences by area of work reveal an expressive gap in \textit{soft skills} training for technical roles.

The central contribution is the evidence that the personal profile of the participant does not determine the perception of quality and effectiveness of training in SE, while professional trajectory produces localized differences and the nature of training (mandatoriness) is the dominant factor. In triangulation with the previous phases of this research \cite{siqueira2026quantitative, siqueira2026qualitative, siqueira2026professional}, which identified instructional design as the central factor, the combined evidence sustains that perceived effectiveness depends more on the participant's professional moment and on how training is offered than on who they are. The absence of gender differences, in particular, constitutes a finding that demands attention: not because women and men perceive training in the same way (which would already be relevant), but because this homogeneity occurs in a context where women are structurally under-represented, suggesting that gender barriers operate before training, in access to the sector.

Future work should prioritize: (1) the inclusion of typological variables differentiating types of training (e.g., \textit{onboarding}, technical updating, \textit{compliance}, \textit{workshops}), allowing the effect of experience to be isolated from the effect of training type; (2) longitudinal studies to assess whether the non-linear experience pattern holds over time; (3) the adoption of multi-item scales with psychometric validation (confirmatory factor analysis, Cronbach's alpha) to overcome the limitation of single items per construct; (4) cross-cultural replications testing whether the homogeneity by gender and education holds in contexts with different diversity profiles; and (5) the exploration of how new technologies, such as LLM-based solutions, can personalize learning paths by career stage, promoting equity and effectiveness \cite{salas2012science, lee2025overcoming}.

\section*{Declaration of the Use of Artificial Intelligence}
This research was originally developed in Portuguese. The authors used Generative AI tools (OpenAI ChatGPT, Google Gemini, and Anthropic Claude) exclusively for support in translating to English, improving textual cohesion and clarity, and assisting with the structuring and revision of the manuscript.

\section*{Artifact Availability}\label{sec:artefatos}
The questionnaire, the question design matrix documenting all four instrument development stages, analysis scripts, and results visualizations are available as open-access artifacts at \href{https://doi.org/10.5281/zenodo.19172188}{Zenodo (10.5281/zenodo.19172188)}.

\bibliographystyle{ACM-Reference-Format}
\bibliography{qualificacao}

\end{document}